\documentclass{statsoc}
\usepackage{geometry}
\usepackage{graphicx,url}
\usepackage[utf8x]{inputenc}

\usepackage{amsmath, amssymb, amsthm, bm,bbm}

\usepackage[english]{babel}
\usepackage{natbib}
\usepackage[colorlinks=true]{hyperref}
\hypersetup{urlcolor=blue, citecolor=blue, linkcolor=blue}
\usepackage{xcolor}

\usepackage{float}
\usepackage{multirow}
\def\bbeta{\bm \beta}
\def\gam{ \gamma}
\def\Sig{\bm \Sigma}
\def\bOmega{\bm \Omega}
\def\tOmega{\widetilde{\bm\Omega}}

\def\bw{\bm w}
\def \by {\bm y}
\def\bx{\bm x}
\def\bX{\bm X}

\def\bb{\bm b}

\def\tSig{\widetilde{\Sig}}
\def\hw{\widehat{\bw}}
\def\tw{\tilde{\bw}}

\def\tn{\tilde{n}}
\def\hS{\widehat{\bm S}}

\def\SNR{\text{SNR}}
\def\tx{\tilde{\bx}}
\def\tX{\tilde{\bm X}}

  \def\sig{\sigma}
  \def\hSig{\widehat{\Sig}}
  \def\R{\mathbb{R}}
  \def\E{\mathbb{E}}
  \def\P{\mathbb{P}}
  
  \def\calH{\mathcal{H}}
  \def\calF{\mathcal{F}}
    
    \def\calZ{\mathcal{Z}}
  \def\eps{\epsilon}
  \def\hbeta{\widehat{\bbeta}}
  
    \def\lam{\lambda}

\DeclareMathOperator*{\argmin}{arg\,min}
\newtheorem{theorem}{Theorem}[section]

\newtheorem{lemma}{Lemma}[section]
\newtheorem{corollary}{Corollary}[section]
\newtheorem{remark}{Remark}[section]
\newtheorem{condition}{Condition}[section]

\title[Inference with proxy data]{Estimation and Inference with Proxy Data and its Genetic Applications}
 
\author[Li {\it et al.}]{Sai Li}
\address{Institute of Statistics and Big Data,
Renmin University of China,
Beijing,
China.}
\author{T. Tony Cai}
	\address{Department of Statistics, The Wharton School, University of Pennsylvania, Philadelphia, PA 19104}
\author[Li, Cai and Li]{Hongzhe Li}
\address{Department of Biostatistics, Epidemiology and Informatics, Perelman School of Medicine, University of Pennsylvania, Philadelphia, PA 19104}
\email{hongzhe@upenn.edu}
\begin{document} 

\maketitle 

\begin{abstract}
Existing high-dimensional statistical methods are largely established for analyzing individual-level data. In this work, we study estimation and inference for high-dimensional linear models where  we only observe  ``proxy data'', which include the marginal statistics and sample covariance matrix that are computed based on different sets of individuals. We develop  a rate optimal method for estimation and inference for the regression coefficient  vector and its linear functionals based on the proxy data. Moreover, we show the intrinsic limitations in the proxy-data based inference: the minimax optimal rate for estimation is slower than that in the conventional case where individual data are observed; the power for testing and multiple testing  does not go to one as the signal strength goes to infinity. 
These interesting findings are illustrated through simulation studies and an analysis of  a dataset concerning the genetic associations of hindlimb muscle weight  in a mouse population.
\end{abstract}

\section{Introduction}
\label{sec1}
 
Large-scale genome-wide association studies (GWAS) provide opportunities for developing
genetic risk prediction models that have the potential to improve disease prevention, intervention, and treatment. 
In epidemiology and genetics, there is a growing interest in utilizing the published summary statistics, especially those from GWAS, for disease risk prediction.  The abundant summary data can enhance the power in signal detection using the framework of meta-analysis. Comparing with the individual-level data, the summary data are less privacy-sensitive  and are more communication efficient for data sharing. However, statistical properties of learning based on the summary data remain largely unknown.

\subsection{Problem formulation}
Let $\bX\in\R^{n\times p}$ denote the standardized genetic variants measurements in $n$ independent individuals whose $i$-th row is $\bx_i^{\intercal}$, where $p$ is the dimension of genetic variants. Let $y\in\R^n$ denote the mean-adjusted response vector in this sample. 
In the linear model for the association between the outcome and the covariates,
\begin{align}
\label{lm1}
   y_i=\bx_i^{\intercal}\bbeta+\eps_i,
\end{align}
where $\E[\eps_i|\bx_i]=0$ and $\E[\eps_i^2|\bx_i]=\sig^2$, the goal is to estimate and infer the effect size vector $\bbeta\in\R^p$ and its functionals using  only the  summary data but not the individual-level data. 
 
GWAS reports the marginal statistics
\[
  \widehat{S}_j=\bX_{.,j}^{\intercal}\by/n,~j=1,\dots, p
\]
and their estimated standard errors.
Besides the marginal statistics $\hS$, an estimator of the covariance matrix of $\bx_i$ is often needed for estimation and inference. One challenge is that the empirical covariance matrix for the samples involved in $\hS$ is often not available because this genomic data set is too large or privacy-sensitive to share. That is, we do not observe $\widehat{\Sig}=\bX^{\intercal}\bX/n$. Let $\Sig$ denote the oracle $\E[\bX^{\intercal}\bX/n]$. A common practice is to obtain  an estimate of covariance matrix $\Sig$ from some external genome panel, such as the 1000 genome project (\url{https://www.internationalgenome.org}).
Let $\tx_i\in\R^p, i=1,\dots,\tn$, denote the genotype of the observations in the external data and define
\[
    \tSig=\frac{1}{\tn}\sum_{i=1}^{\tn}\tx_i\tx_i^{\intercal}.
\]
We call $\tx_i, ~i=1,\dots,\tn$, the proxy data and $\tSig$ the proxy covariance matrix.
In this work, we assume that $\E[\tx_i\tx_i^{\intercal}]=\Sig$. That is, the proxy covariates have the same covariance structure as the covariates in computing the marginal statistics $\hS$.  
In genetic applications, the number of SNPs can be much larger than the sample sizes. Hence, we develop methods for the regime that  $p$ is larger or much larger than $\max\{n,\tn\}$.

\subsection{Motivating applications with the proxy data}
Learning the linear model  (\ref{lm1}) with summary statistics  has important applications in genomic studies.
Polygenic risk score (PRS) regression concerns predicting a certain health-related outcome using the associated single nucleotide polymorphisms (SNPs). In fact, PRS can be formulated as a high-dimensional regression problem \citep{Chen20}. It is crucial in PRS prediction to provide confidence intervals for $\bx_*^{\intercal}\bbeta$ given a new individual's genomic information $\bx_*\in\R^p$. 
Besides, summary data provide the opportunity to combine multiple studies (e.g., cohorts) into one large study to increase the sample size \citep{albinana2020leveraging}, which is also the goal of meta-analysis \citep{deelen2019meta}. Hence, it is also of statistical interest to estimate and make inference of $\bbeta$ with the proxy data.

Another  application of the proxy-data based inference is distributed inference, where the whole data set contains \textit{i.i.d.} observations but the data are distributed at multiple remote machines. Distributed algorithms estimate the target parameter by communicating some summary information across machines. To reduce the communication costs, the gradient vectors are communicated but not the high-dimensional Hessian matrix. The overall Hessian matrix can be approximated by a local matrix or by subsampling. See, for example, \citet{jordan2018communication} and \citet{wang2019utilizing}. This type of distributed inference also falls in the category of proxy-data based inference. In Section \ref{sec-diss}, we discuss some other modern applications of the proxy-data based inference in causal inference and genetics.

To summarize, estimation and inference for $\bbeta$ and $\bx_*^{\intercal}\bbeta$ based on the proxy data $\hS$ and $\tSig$ have significant practical values. The statistics $\hS$ and $\tSig$ are also known as two-sample summary data. 
 In the existing literature, the ``two-sample'' setting has been largely used to refer to having two samples from different distributions, such as two-sample testing. To avoid confusion, the term ``proxy data'' is adopted. 

Motivated by aforementioned applications, we study proxy-data based statistical inference  in high-dimensional linear models. The key challenge is that the sample covariance matrix observed $\tSig$ is exclusive of the samples for computing $\hS$.  This distinguishes the current problems with, say, the semi-supervised problems and transfer learning problems. We highlight some of our key contributions.

\subsection{Main results and our contribution}
\label{sec1-3}


Methodology-wise, we consider a proxy-data based $\ell_1$-penalized regression estimator and prove that it is minimax optimal under typical regularity conditions. We further propose debiased estimators to make inference for $\beta_j$ for some fixed $1\leq j\leq p$ and $\bx_*^{\intercal}\bbeta$ with the proxy data, respectively. The debiased estimators are asymptotically normal and can be used to construct confidence intervals and for multiple testing under certain conditions. We also demonstrate that confidence interval length given by the debiased estimator of $\beta_j$ has minimax optimal length under certain conditions.
 
 Theoretically, we discover some interesting and new phenomena with the proxy-data based learning. First, the minimax rates for estimation are slower than the corresponding rates with one-sample individual data, even if $\tn\rightarrow \infty$. The relative loss gets larger when the signal-to-noise ratio gets larger. 
Second, comparing with the debiased Lasso based on individual data, the debiased Lasso estimator of $\beta_j$ based on proxy data has larger bias and variance and its asymptotic normality requires a different sample size conditions (Theorem \ref{thm-db}).  In addition, the results also imply that simply treating the two-sample summary data as one-sample data can lead to invalid inference.
Third, the power of the proxy data-based inference is always no larger than the power in the conventional setting  even if $\tn\rightarrow \infty$. In fact, the power function based on the proxy data is upper truncated by a function related to the magnitude of $\min\{n, \tn\}/s$.  This demonstrates a curse for the proxy data-based inference with dense signals: for finite $\min\{n, \tn\}/s$, the power for strong signals is largely deminished.

%

\subsection{Related literature}
Estimation and inference for the regression coefficients have been extensively studied in high-dimensional linear models based on individual data from one sample. Many penalized methods have been proposed for prediction, estimation, and variable selection in high-dimensional linear models. To name a few, \citet{Lasso, FL01, Zou06, CT07, MB10, Zhang10}. 
Statistical inference for each regression coefficient has been studied in the conventional setting. One stream of methods is inference based on consistent variable selection. Under the assumption that the minimal signal strength is sufficiently large \citep{zhao2006model,wainwright2009sharp}, all the true signals can be consistently selected based on regularization \citep{Lasso, FL01, Zhang10}. Hence, the high-dimensional model is reduced to the low-dimensional problem and classical fixed-dimensional inference tools can be applied. 
The second stream of methods does not rely on the minimal signal strength condition.
\citet{ZZ14}, \citet{van14}, and \citet{JM14} consider debiased estimators in linear models and generalized linear models. The minimaxity and adaptivity of confidence intervals have been studied in \citet{CG17}. \citet{NL17} proposes a general framework to de-bias regularized estimators in different models.  
Weaker sample size conditions or sparsity conditions have been studied in \citet{JM19} and \citet{ZB18} under certain assumptions. \citet{cai2019individualized} and \citet{javanmard2020flexible} propose methods for inference for a linear functional of the regression coefficients in linear models.

The proxy data-based estimation and prediction methods have appeared in genetic applications. \citet{LDpred} introduces an Bayesian approach for PRS based on summary data. \citet{Lassosum} and \citet{Chen20} both consider shrinkage methods as extensions of the Lasso. The method in \citet{Lassosum} is for linear models and the method in \citet{Chen20} can deal with binary outcomes based on approximations. However, the statistical guarantees and the choice of tuning parameters are largely unknown. In addition, 
inference for the linear functionals of high-dimensional regression coefficients based on proxy data has not been studied in literature.

\subsection{Organization and notation}
In Section \ref{sec-method}, we describe the $\ell_1$-regularized method for estimating $\bbeta$ with high-dimensional proxy data and study its convergence rate and minimax optimality. In Section \ref{sec3}, we construct the debiased estimator of $\beta_j$ based on proxy data and study its limiting distribution 
In Section \ref{sec3-3}, we construct confidence interval for $\bx_*^{\intercal}\bbeta$ and provide theoretical guarantees. In Section \ref{sec-simu}, we study the empirical performance of our proposals via extensive numerical experiments. In Section \ref{sec-data}, we apply the one-sample and two-sample methods to a GWAS in an outbred mice population. In Section \ref{sec-diss}, we discuss some other summary data motivated problems for future research. The proofs and other supplementary information are provided in the supplements \citep{Supp}.

\textbf{Notation}. For real-valued sequences $\{a_n\}, \{b_n\}$, we write $a_n \lesssim b_n$ if $a_n \leq cb_n$ for some universal constant $c \in (0, \infty)$, and $a_n \gtrsim b_n$ if $a_n \geq c'b_n$ for some universal constant $c' \in (0, \infty)$.  We say $a_n \asymp b_n$ if $a_n \lesssim b_n$ and $a_n \gtrsim b_n$. $c, C, c_0, c_1, c_2, \cdots, $ and so on refer to universal constants in the paper, with their specific values possibly varying from place to place. 
For a vector $\bm v \in \R^d$ and a subset $S \subseteq [d]$, we use $\bm v_S$ to denote the restriction of vector $\bm v$ to the index set $S$. For a matrix $A\in\R^{n_1\times n_2}$, let $\Lambda_{\max} (A)$ denote the largest singular value of $A$, $\Lambda_{\min}(A)$ denote the smallest singular value of $A$, and $\|A\|_{\infty,\infty}$ denote $\max_{i\leq n_1,j\leq n_2}|A_{i,j}|$. For a random variable $u\in\R$, define its sub-Gaussian norm as $\|u\|_{\psi_2}=\sup_{l\geq 1} l^{-1/2}\E^{1/l}[|u|^l]$. For a random vector $\bm U\in\R^{n}$, define its sub-Gaussian norm as $
   \|\bm U\|_{\psi_2}=\sup_{\|\bm v\|_2=1,\bm v\in\R^{n}}\|\langle \bm U,\bm v\rangle\|_{\psi_2}$. 
Let $\text{SNR}=\|\Sig^{1/2}\bbeta\|_2^2/\sig^2$ denote the signal-to-noise ratio. Let $\tau_q$ denote the $q$-th quantile of standard normal distribution.

\section{Estimation and prediction based on proxy data}
\label{sec-method}
In this section, we introduce our proposed estimators for prediction and estimation based on proxy data in Section \ref{sec2-1}. We study its theoretical properties and minimax optimality in Section \ref{sec2-2}.

\subsection{Two-sample Lasso method}
\label{sec2-1}
For the estimation and prediction tasks, the methods for proxy data resemble one-sample high-dimensional regression methods. The Lasso estimator \citep{Lasso} provides 
a rate optimal estimator of $\bbeta$ in the conventional setting. Decomposing the empirical loss $\|\by-\bX{\bm b}\|_2^2$ as $\|\by\|_2^2-2{\bm b}^{\intercal}\bX^{\intercal}\by+\|\bX{\bm b}\|_2^2$ and removing the constant term, the Lasso estimator can be written as
\[
  \hbeta^{(os)}=\argmin_{\bm b\in\R^p}\left\{\frac{1}{2}{\bm b}^{\intercal}\hSig {\bm b}-{\bm b}^{\intercal}\hS+\lam^{(os)}\|{\bm b}\|_1\right\}
\]
with some tuning parameter $\lam^{(os)}>0$ and the superscript ``os'' is short for ``one-sample''.  In fact, we have seen that the Lasso can be equivalently performed based on one-sample summary data $\hSig$ and $\hS$. Methodology-wise, there is no need to distinguish one-sample summary and individual data for the Lasso. In the sequel, we will refer to $\hbeta^{(os)}$ as one-sample Lasso for simplicity.

With proxy data, it is natural to consider the following estimatior
\begin{align}
\label{eq-cbeta}
  \hbeta^{(ts)}=\argmin_{\bm b\in\R^p}\{\frac{1}{2}{\bm b}^{\intercal}\tSig {\bm b}-{\bm b}^{\intercal}\hS+\lam^{(ts)}\|{\bm b}\|_1\},
\end{align}
where we replace the unknown $\hSig$ with its proxy $\tSig$ and consider a possibly different tuning parameter $\lam^{(ts)}$.
We will see later that $\lam^{(ts)}$ should always be chosen larger than $\lam^{(os)}$ for consistency. 

The two-sample Lasso estimator $\hbeta^{(ts)}$ has been considered in \citet{Lassosum} and \citet{Chen20}. However, the choice of $\lam^{(ts)}$, the convergence rate, and minimax optimality have not been established. Choosing $\lam^{(ts)}$ is also a practical challenge because cross validation cannot be performed  without individual-level data. We provide the theoretical requirement on $\lam^{(ts)}$ in Section \ref{sec2-2} and discuss some practical choices based on information criteria in Section \ref{sec-simu}.

\subsection{Convergence rates for estimation and prediction}
\label{sec2-2}
We assume the following conditions for theoretical analysis.
\begin{condition}[Gaussian designs]
\label{cond1}
Each row of $\bX$ and $\tX$  are i.i.d. Gaussian with mean zero and positive definite covariance $\Sig$ with bounded eigenvalues. 
\end{condition}

\begin{condition}[sub-Gaussian noises]
\label{cond2}
The random noises $\eps_i$, $i=1,\dots,n$, are i.i.d. with mean zero and variance $\sig^2>0$. $\eps_i$ and $\bx_i$ are independent for $i=1,\dots,n$. The sub-Gaussian norms of $\eps_i$ are upper bounded by a constant.
\end{condition}
For estimation and prediction, it suffices to relax Conditions \ref{cond1} and \ref{cond2}  to assume independent sub-Gaussian designs and independent sub-Gaussian noises. Here we assume slightly stronger regularity conditions, Gaussian designs and \textit{i.i.d.} noises. These assumptions ensure that the asymptotic variance of the debiased estimators only depends on the first two moments of the observations. With individual samples, this assumption is not necessary because one can estimate the variance based on the empirical noises \citep{Dezeure17}. In lack of the individual-level data, we cannot estimate the asymptotic variance empirically and have to rely on the properties of higher moments, which makes Gaussian distribution a natural  assumption.

We first  derive the rate of convergence for $\hbeta^{(ts)}$ in the two-sample summary setting. Let $\E[y_i^2]=M$ and
\begin{align}
\label{eq-gam}
    \gam_{n,\tn}=\sig^2+\|\Sig^{1/2}\bbeta\|_2^2(\frac{n}{\tn}+1)=M+\frac{n}{\tn}\bbeta^{\intercal}\Sig\bbeta.
\end{align}
Loosely speaking, $\gam_{n,\tn}$ is the variance of the random noises based on proxy data. In Theorem \ref{lem2-est}, we establish the convergence rate of $\hbeta^{(ts)}$ under mild conditions.
\begin{theorem}[Convergence rates for $\hbeta^{(ts)}$]
\label{lem2-est}
Assume Conditions \ref{cond1} and \ref{cond2} and $Ms\log p\ll \min\{n,\tn\}$.
For $\lam^{(ts)}= c_1\sqrt{
\gam_{n,\tn}\log p/n}$, with large enough $c_1,c_2$,  then with probability at least $1-\exp(-c_1\log p)-\exp(-c_2\tn)$,
\begin{align*}
&\|\tSig^{1/2}(\hbeta^{(ts)}-\bbeta)\|_2^2\vee \|\hbeta^{(ts)}-\bbeta\|_2^2\leq C\frac{\gam_{n,\tn}s\log p}{n}\\
 & \|\hbeta^{(ts)}-\bbeta\|_1\leq Cs\sqrt{\frac{\gam_{n,\tn}\log p}{n}}.
\end{align*}
\end{theorem}
Comparing with the one-sample optimal rates in squared $\ell_2$-norm, which is $\sig^2s\log p/n$, we can see that the ratio of two rates (two-sample over one-sample) is
\begin{align}
\label{ratio1}
      1+\SNR(\frac{n}{\tn}+1),
\end{align}
where $\text{SNR}=\|\Sig^{1/2}\bbeta\|_2^2/\sig^2$.
This implies that the estimation error rate in two-sample case is strictly worse than the one-sample case as long as $\SNR> 0$. Larger $n$ leads to  larger  relative loss with $\hbeta^{(ts)}$ relative to $\hbeta^{(os)}$. In constrast, larger $\tn$ implies  smaller  relative loss. 
Finally, a larger  SNR implies a larger loss of $\hbeta^{(ts)}$ compared to $\hbeta^{(os)}$.  
As a result, the condition for consistency is no weaker in the proxy setting than that with one-sample data.
The two-sample tuning parameter $\lam^{(ts)}\asymp \sqrt{\gam_{n,\tn}\log p/n}$, whose order is always no smaller than its one-sample counterpart. The choice of $\lam^{(ts)}$ is crucial in practice and we will discuss this in Section \ref{sec-simu}.

To better understand the unique challenges with proxy data, we consider a special scenario where $\tn\rightarrow \infty$, or equivalently, $\Sig$ is known. 
  \begin{remark}[The scenario of $\tn\rightarrow \infty$]
  \label{re3-1}
  {\rm
If $\tn\rightarrow \infty$, which is equivalent to observing $(\hS,\Sig)$, then 
\[\|\hbeta^{(ts)}-\bbeta\|_2^2=O_P\left(\frac{Ms\log p}{n}\right).
\]
 } \end{remark} 
Remark \ref{re3-1} shows that even if  $\tn\rightarrow \infty$, the convergence rate of $\hbeta^{(ts)}$ is still inflated when $\SNR> 0$ in comparison to having one-sample data. This comparison implies that, without the in-sample $\widehat{\Sig}$, any estimator of $\Sig$, even the oracle one, can lead to dramatic loss in estimation accuracy. Comparing Remark \ref{re3-1} with Theorem \ref{lem2-est}, we see that the error caused by finite external data is  of order $\bbeta^T\Sig\bbeta s\log p/\tn$.


We now show that the convergence rate of $\hbeta^{(ts)}$ is minimax optimal in $\ell_2$-norm. 
Consider  the parameter space
 \begin{align}
  \Xi(s,M_0,\sig_0^2)&=\left\{\|\bbeta\|_0\leq s, \bbeta^{\intercal}\Sig\bbeta\leq M_0, ~0<\sig^2\leq \sig_0^2,\right.\nonumber\\
  &\quad~~\left. 0<1/C_1\leq \Lambda_{\min}(\Sig)\leq \Lambda_{\max}(\Sig)\leq C_1<\infty\right\} \label{eq-Xi}
\end{align}
for some constant $C_1>1$ and $\sig_0^2$ can be any positive constant. We see that  $M\leq M_0+\sig_0^2$ in the space of $  \Xi(s,M_0,\sig_0^2)$.
Let $\calZ=\{\hS,\tSig\}$ denote the observed data and $\mathcal{F}(\calZ)$ denote functions based on the summary data $\calZ$.

\begin{theorem}[Lower bound for estimating $\bbeta$]
\label{thm-mini-est} 
Consider the parameter space $\Xi(s,M_0,\sig_0^2)$ in (\ref{eq-Xi}) with $s\geq 2$.
Suppose that $Ms\log p\ll n$, and $(\bbeta^T\Sig\bbeta\vee 1)s\log p\ll \tn$.
Then there exists some constant $c_2$ that
\[
  \min_{\hbeta\in\calF(\calZ)}\sup_{\bbeta\in\Xi(s,M_0,\sig_0^2)}\P\left(\|\hbeta-\bbeta\|_2^2\geq \frac{c_1(M_0+\sig_0^2)s\log p}{n}+\frac{c_2M_0s\log p}{\tn}\right)\geq 1/2.
\]
\end{theorem}
In the parameter space $\Xi(s,M_0,\sig_0^2)$, it holds that $M=\E[y_i^2]\leq M_0+\sig_0^2$. Hence, the lower bound in Theorem \ref{thm-mini-est} matches the $\ell_2$-upper bound in Theorem \ref{lem2-est} in terms of rates. We mention that the sample size condition in Theorem \ref{thm-mini-est} essentially restricts us to a class of $\hS$ with distributional regularity, i.e., its distribution conditioning on $\by$ has positive definite covariance matrix. As far as we know, this is the first lower bound result based on summary data and the proof is based on some novel analysis of the distribution of the marginal correlation statistics.

\section{Inference for individual coefficient based on proxy data}
\label{sec3}
In this section, we consider statistical inference, such as hypothesis testing and constructing confidence intervals for $\beta_j$ with some fixed $1\leq j\leq p$. 
It is known that the $\ell_1$-regularized estimates are biased and cannot be directly used for inference. 
For inference based on proxy data, we follow a similar idea as the debiased methods, which have been proposed based on one-sample individual data. Specifically, the debiased Lasso \citep{ZZ14, van14, JM14} can be written as
\begin{align}
\label{os-db}
   \hat{\beta}_j^{(os-db)}&=\hat{\beta}^{(os)}_j+\frac{(\bX\hw_j)^{\intercal}(\by-\bX\hbeta^{(os)})}{n}=\hat{\beta}^{(os)}_j+\hw_j^{\intercal}(\hS-\widehat{\Sig}\hbeta^{(os)}),
\end{align}
where $\hbeta^{(os)}$ is the one-sample Lasso estimator and $\hw_j\in\R^p$ is a correction score vector that  can be computed based on $\widehat{\Sig}$. We see that the debiased Lasso for $\beta_j$ can also be realized based on the summary statistics $\hSig$ and $\hS$. Hence, we refer to the estimate in (\ref{os-db}) as one-sample debiased Lasso (os-db) in the sequel.
This similarly motivates its counterpart with two-sample summary data:
 \begin{align}
 \label{ts-db}
   \hat{\beta}_j^{(ts-db)}&=\hat{\beta}^{(ts)}_j+\tw_j^{\intercal}(\hS-\tSig \hbeta^{(ts)}),
\end{align}
where $\hbeta^{(ts)}$ is computed in (\ref{eq-cbeta}) and $\tw_j\in\R^p$ is a correction score vector computed based on $\tSig $.
Specifically, we consider 
\begin{align}
 \tw_j&=\argmin_{\bw\in\R^p}\|\bw\|_1\label{eq-tw}\\
 &\text{subject to}~\|\tSig\bw-{\bm e}_j\|_{\infty}\leq \lam_j,\nonumber
\end{align}
where $\lam_j=c_1\sqrt{\log p/\tn}$ for some positive constant $c_1$.
The realization of $\tw_j$ is via a Dantzig selector optimization \citep{CT07}, which induces a sparse solution of the $j$-th column of the inverse covariance matrix $\bOmega=\Sig^{-1}$. Some existing one-sample methods, such as \citet{JM14}, do not look for a sparse estimate $\tw_j$ but choose a different objective function in (\ref{eq-tw}). In the proxy setting, however, the sparsity of $\tw_j$ plays a crucial role in the analysis. Those non-sparse methods for one-sample setting cannot be directly generalize for the current purpose as we will further discuss in Section \ref{sec3-2}.

\subsection{Asymptotic normality for debiased two-sample Lasso}
\label{sec3-2}
We study the asymptotic property of $\hat{\beta}_j^{(ts-db)}$ defined in (\ref{ts-db}) and prove its asymptotic normality under certain conditions.
Let $\bOmega_{.,j}$ denote the $j$-th column of $\bOmega$ and $s_j=\|\bOmega_{.,j}\|_0$. 
    \begin{theorem}[Asymptotic normality of the debiased estimator]
    \label{thm-db}
 Assume that Condition \ref{cond1} and Condition \ref{cond2} hold, $n\gg \log p$, and $\tn\gg (s\vee s_j)\log p$. Then it holds that
    \begin{equation}
    \label{re1-thm-db}
       \hat{\beta}^{(ts-db)}_j-\beta_j=z_j+O_P\left(\gam_{n,\tn}^{1/2}\frac{(s+s_j)\log p}{\sqrt{n\tn}}\right),
    \end{equation}
    where $\gam_{n,\tn}$ is defined in (\ref{eq-gam}) and
    \[
       (V_j^{(ts)})^{-1/2}z_j\xrightarrow{D} N(0,1)~~ \text{for}~~ V_j^{(ts)}=\frac{\Omega_{j,j}\gam_{n,\tn}}{n}+\frac{\beta_j^2}{n}+\frac{\beta_j^2}{\tn}.
       \]
Further assuming $( s\vee s_j)\log p \ll \sqrt{\tn}$, 
  then $(V_j^{(ts)})^{-1/2}(\hat{\beta}^{(ts-db)}_j-\beta_j)\xrightarrow{D} N(0,1)$. 
    \end{theorem}

Theorem \ref{thm-db} establishes the asymptotic distribution of $\hat{\beta}_j^{(ts-db)}$ in (\ref{re1-thm-db}) and provides the sample size condition for its asymptotic normality. The variance of $\hat{\beta}_j^{(ts-db)}$ is $V_j^{(ts)}$ and the remaining bias of $\hat{\beta}^{(ts-db)}_j$ is shown in the last term on the right hand side of (\ref{re1-thm-db}). 

We first bring some details into the magnitude of $V_j^{(ts)}$.
The asymptotically normal component is $z_j=\bOmega_{.,j}^{\intercal}(\hS-\Sig\bbeta)+\bOmega_{.,j}^{\intercal}(\Sig-\tSig)\bbeta$, where the first term comes from the marginal statistics and the second term comes from the proxy matrix. The variance $V_j^{(ts)}$ is obtained based on the moment formula for multivariate Gaussian.
The last two terms of $V_j^{(ts)}$, $\beta_j^2/n$ and $\beta_j^2/\tn$, are dominated by the first term of $V_j^{(ts)}$ given the positive definiteness of $\bOmega$.
Hence, when $\tn\gg n$, $V_j^{(ts)}\asymp \Omega_{j,j}M/n$; when $n\gg \tn$, $V_j^{(ts)}\asymp \Omega_{j,j}\|\Sig^{1/2}\bbeta\|_2^2/\tn$.  In comparison to its one-sample counterpart, $V_j^{(os)}=\Omega_{j,j}\sig^2/n$, the relative loss in efficiency is
\[
  \frac{V_j^{(ts)}}{V_j^{(os)}}\asymp 1+\SNR(\frac{n}{\tn}+1),
\]
which is identical to the relative loss in estimation (\ref{ratio1}). When $\tn\rightarrow \infty$, i.e. $\Sig$ is known, $V_j^{(ts)}$ is still larger than $V_j^{(os)}$. This shows the significant loss in efficiency for inference problems when the marginal statistics and covariance estimator are not based on the same set of samples.
More importantly, the distinction between $V_j^{(ts)}$ and $V_j^{(os)}$ implies that simply applying the one-sample inference algorithms to the two-sample data could be wrong. We illustrate this point numerically in Section \ref{sec-simu-3}. 

The remaining bias of $\hat{\beta}^{(ts-db)}_j$ is controlled by the sample sizes $\sqrt{n\tn}$ and $\tn$. This implies that the sample size for external reference panel, $\tn$ plays a more significant role in controlling the bias while  the sample size for GWAS, $n$, plays a milder role. We have seen the same phenomenon in the estimation results in Section \ref{sec2-2}.
In one-sample setting, the remaining bias of debiased $\hat{\beta}_j^{(os-db)}$ in (\ref{os-db}), is of order $s\log p/n$ \citep{van14, CG17}. If $\tn/n$ is sufficiently large, the remaining bias of $\hat{\beta}^{(ts-db)}_j$ can be \textbf{smaller} than that of its one-sample counterpart. In view of the asymptotic bias and asymptotic variance in (\ref{re1-thm-db}), it suffices to require $(s\vee s_j)\log p\ll \sqrt{\tn}$ for asymptotic normality. This condition implies that $\tn$ determines the range of sparsity such that valid inference can be established. In contrast, for one-sample debiased Lasso $\hat{\beta}_j^{(os-db)}$, $n$  determines the range of sparsity for valid inference, which is $s\log p\ll \sqrt{n}$. This can be a blessing of proxy-data scenario. An extreme case is when $\Sig$ is known, or equivalently $\tn\rightarrow \infty$ as in the following remark.
\begin{remark}[The scenario of $\tn\rightarrow \infty$]
  \label{re1-inf}
  {\rm
When $\Sig$ is known, $\tilde{\bw}_j=\bOmega_{.,j}$ and $\hat{\beta}_j^{(ts-db)}=\bOmega_{j,.}\bX^{\intercal}\by/n$, which is asymptotically normal with mean zero and variance $\Omega_{j,j}\gam_{n,\infty}/n+\beta_j^2/n$.
 } \end{remark} 
It may be surprising to see that for fixed $n$ and $p$,  the remaining bias of $\hat{\beta}^{(ts-db)}_j$ vanishes when $\tn\rightarrow \infty$. However, many existing applications often have $\tn\lesssim n$. This can be due to the less cost of sharing GWAS statistics than sharing the LD matrix. Same pattern holds for distributed inference, in which case $n$ is the total sample size and $\tn$ is the local sample size. This should raise some caution in applications with two-sample summary data.

We finally  discuss the conditions on the sparsity $s_j$. In classical one-sample setting, inference for $\beta_j$ may not require sparse $\bOmega_{.,j}$, see, for example, the analysis in \citet{JM14} for linear models. We mention that the condition on $s_j$ cannot be removed using the same idea in our anlaysis. This comes from a unique challenge of proxy data, where $\hS$ implicitly depends on $\hSig$, which is unobserved but approximated. Nevertheless, the condition on $s_j$ can be avoided by sample splitting. Ideally, one can create two independent estimate of $\Sig$ and use one for two-sample Lasso and the other one as the debiasing samples.  
However, sample splitting is not viable with summary data in most cases. Hence, we focus on the current procedure and the results without sample splitting.

In the next theorem, we establish the minimax lower bound for estimating $\beta_j$. 
\begin{theorem}[Minimax lower bound for estimation of $\beta_j$]
\label{thm-mini-inf}
Consider the parameter space $\Xi(s,M_0,\sig_0^2)$ in (\ref{eq-Xi}).
Suppose that $\max\{1,M_0+\sig_0^2\}\leq c_1\min\{n, \tn\}$ for some constant $c_1>0$. Then for any fixed $1\leq j\leq p$,  there exists some constant $c_2$ that
\[
   \inf_{\hat{\beta}_j\in\calF(\calZ)} \sup_{\bbeta\in\Xi(s,M_0,\sig_0^2) }\P\left(|\hat{\beta}_j-\beta_j|\geq c_2\sqrt{\frac{M_0+\sig_0^2}{n}}+c_2\sqrt{\frac{M_0}{\tn}}\right)\geq \frac{1}{2}.
\]
\end{theorem}
In Theorem \ref{thm-mini-inf}, we show that the parametric part of the rate for $\hat{\beta}_j^{(ts-db)}$ is minimax optimal. That is, under the sample size condition $(s\vee s_j)\log p\ll \sqrt{\tn}$, the two-sample debiased estimator $\hat{\beta}_j^{(ts-db)}$ has rate optimal confidence interval length.
Comparing with the minimax rate for one-sample inference, we see that the variance part are inflated with proxy data. For the nonparametric part, the proof based on summary statistics is much more involved. 
In the supplements, we provide the minimax lower bound for estimating $\beta_j$ when $\Sig$ is known and the lower bound matches the upper bound derived in Remark \ref{re1-inf}.

\subsection{Variance estimator and confidence intervals}
In view of $V_j^{(ts)}$, we propose a variance estimator for $\hat{\beta}_j^{(ts-db)}$ as
\begin{align}
   \widehat{V}^{(ts)}_j&=\tw_j^{\intercal}\tSig\tw_j(\frac{\|\by\|_2^2}{n^2}+\frac{2(\hbeta^{(ts)})^{\intercal}\hS-(\hbeta^{(ts)})^{\intercal}\tSig\hbeta^{(ts)}}{\tn})\nonumber\\
   &\quad+\frac{(\hat{\beta}_j^{(ts-db)})^2}{n}+\frac{(\hat{\beta}_j^{(ts-db)})^2}{\tn}.\label{eq-cVj}
\end{align}
Notice that $\widehat{V}^{(ts)}_j$ is \textbf{not} the two-sample analogy of variance estimator for the classical debiased Lasso. This is because the probabilistic limit of $\widehat{V}^{(ts)}_j$ is asymptotically larger than the asymptotic variance of the conventional debased Lasso. Hence, if we treat proxy data as one-sample summary data, correct coverages are not guaranteed. We propose the following $(1-\alpha)\times 100\% $-confidence interval for $\beta_j$ as
\begin{equation}
\label{ci}
    \hat{\beta}_j^{(ts-db)}\pm \tau_{\alpha/2}\sqrt{\widehat{V}_j^{(ts)}}.
\end{equation}
Once the $z$-statistics $ z_j^{(ts)}=\hat{\beta}_j^{(ts-db)}/\sqrt{\widehat{V}_j^{(ts)}}$ is obtained for $j=1,\dots, p$, we can perform multiple testing with FDR control using the procedure in \citet{JJ19}, which is a refined version based on \citet{Liu13}.

In the next lemma, we prove the consistency of $\widehat{V}_j^{(ts)}$ defined in (\ref{eq-cVj}) and conclude the validness of the confidence interval (\ref{ci}).
\begin{lemma}[A consistent variance estimator]
\label{lem-var}
Under the conditions of Theorem \ref{thm-db}, 
\[
  \frac{|\widehat{V}_j^{(ts)}-V_j^{(ts)}|}{V_j^{(ts)}}=O_P\left(\frac{\gam_{n,\tn}}{\sqrt{n}}+\gam_{n,\tn}\frac{s\log p}{n}+\gam_{n,\tn}\frac{s_j\log p}{\tn}+\frac{1}{\sqrt{\tn}}\right).
\]
To summarize, assuming  Condition \ref{cond1}, Condition \ref{cond2}, and $(s\vee s_j)\log p\ll \sqrt{\tn}$, then  $(\widehat{V}_j^{(ts)})^{-1/2}(\hat{\beta}^{(ts-db)}_j-\beta_j)\xrightarrow{D} N(0,1)$. 
\end{lemma}
So far we have proved the asymptotic validness of the confidence interval in (\ref{ci}) under the conditions of Theorem \ref{thm-db}.

\subsection{Power analysis with proxy data}
\label{sec-st}
In this section, we evaluate the power of hypothesis testing with proxy data. We have seen in Section \ref{sec-method} that it is necessary to focus on the regime that $s\log p\ll \sqrt{\tn}$ and $s_j\log p\ll \tn/\sqrt{n}+\tn^{1/2}$ for valid inference.
For the simplicity of the power analysis, we ignore the asymptotic bias in the debiased estimator.
To avoid confusion, we introduce some new notations.
Let $\sqrt{n}\hat{z}^{(os)}_j$ be the probabilistic limit of conventional debiased estimator $\sqrt{n}\hat{\beta}_j^{(db)}/\sqrt{V_j}$ where $V_j=\Omega_{j,j}\sig^2$ \citep{van14}. Let $\bm\Omega=\Sig^{-1}$. The distribution of one-sample $z$-score is
\begin{equation}
\label{z-os}
   \sqrt{n}\hat{z}^{(os)}_j\sim N\left(\frac{\sqrt{n}\beta_j}{\sqrt{\Omega_{j,j}\sig^2}},1\right),~1\leq j\leq p.
\end{equation}

For proxy-data based inference, let $\sqrt{n}\hat{z}_j^{(ts)}$ be the probabilistic limit of two-sample debiased estimator $\sqrt{n}\hat{\beta}^{(ts-db)}_j/\sqrt{V_j^{(ts)}}$. The marginal distribution of each two-sample $z$-score is
\begin{equation}
\label{z-ts}
 \sqrt{n} \hat{z}^{(ts)}_j\sim N\left(\frac{\sqrt{n}\beta_j}{\sqrt{\Omega_{j,j}\gam_{n,\tn}+\beta_j^2(1+n/\tn)}},1\right),~1\leq j\leq p.
\end{equation}
We see that the distribution of $\hat{z}^{(ts)}_j$ depends not only on $\beta_j$ but all other coefficients $\beta_{-j}$.

We evaluate the power of single hypothesis testing, $\calH_{0,j}:~\beta_j=0$ vs $\calH_{1,j}:~\beta_j\neq 0$ based on $\hat{z}_j^{(os)}$ and $\hat{z}_j^{(ts)}$, respectively. By definition, the power of these two statistics can be expressed as
\begin{align*}
 \text{Power}^{(os)}_j(\alpha;\bb)=\P(|\hat{z}_j^{(os)}|\geq \tau_{\alpha}|\beta_j=b_j,\bbeta_{-j}=\bb_{-j}).\\
  \text{Power}^{(ts)}_j(\alpha,\bb)=\P(|\hat{z}_j^{(ts)}|\geq \tau_{\alpha}|\beta_j=b_j,\bbeta_{-j}=\bb_{-j}).
\end{align*}
Based on (\ref{z-os}), we know that $\text{Power}^{(os)}_j(\alpha;\bb)$ is independent of $\bb_{-j}$.

\begin{theorem}[Power of two-sided single hypothesis testing]
\label{thm1-st}
For any $\bb\in\R^p$, it holds that
{\small
\begin{align*}
&\textup{Power}^{(os)}_j(\alpha,\bb)=\Phi(|\eta_j^{(os)}|-\tau_{\alpha})+\Phi(-|\eta_j^{(os)}|-\tau_{\alpha}),~\text{where}~\eta_j^{(os)}=\frac{\sqrt{n}b_j}{\Omega_{j,j}^{1/2}\sig}.\\
&\textup{Power}^{(ts)}_j(\alpha,\bb)=\Phi\left(|\eta_j^{(ts)}|-\tau_{\alpha}\right)+\Phi\left(-|\eta_j^{(ts)}|-\tau_{\alpha}\right),~\text{where}~\\
&\quad\eta_j^{(ts)}=\frac{\sqrt{n}b_j}{\sqrt{(\Omega_{j,j}\bb^{\intercal}\Sig\bb+b_j^2)(1+n/\tn)+\Omega_{j,j}\sig^2}}.
\end{align*}
}
\end{theorem}

The one-sample and two-sample power functions are increasing functions of $|\eta_j^{(os)}|$ and $|\eta_j^{(ts)}|$, respectively.
The power based on proxy data is smaller than that based on one-sample data for $\beta_j\neq 0$. Furthermore, the power based on $\hat{z}^{(os)}_j$ is independent of effect size distribution but  the power based on $\hat{z}^{(ts)}_j$ depends on the effect size distribution. Specifically, if the effect size vector $\bb$ has sparsity $s$ and approximately equal signal strength, then $\eta_j^{(ts)}$ is approximately $\sqrt{n}b_j/\sqrt{\Omega_{j,j}\{sb_j^2(1+n/\tn)+\sig^2\}}$. We see that $|\eta_j^{(ts)}|$ is bounded away from infinity for finite $n\wedge \tn$ even if $b_j\rightarrow\infty$.

To further demonstrate this phenomenon, we consider the equal signal strength model
\begin{equation}
\label{em}
   b_j\in\{-b_0,0,b_0\}~\text{and}~\|\bb\|_0=s.
\end{equation}

\begin{figure}[H]
\includegraphics[width=0.495\textwidth,height=5.7cm]{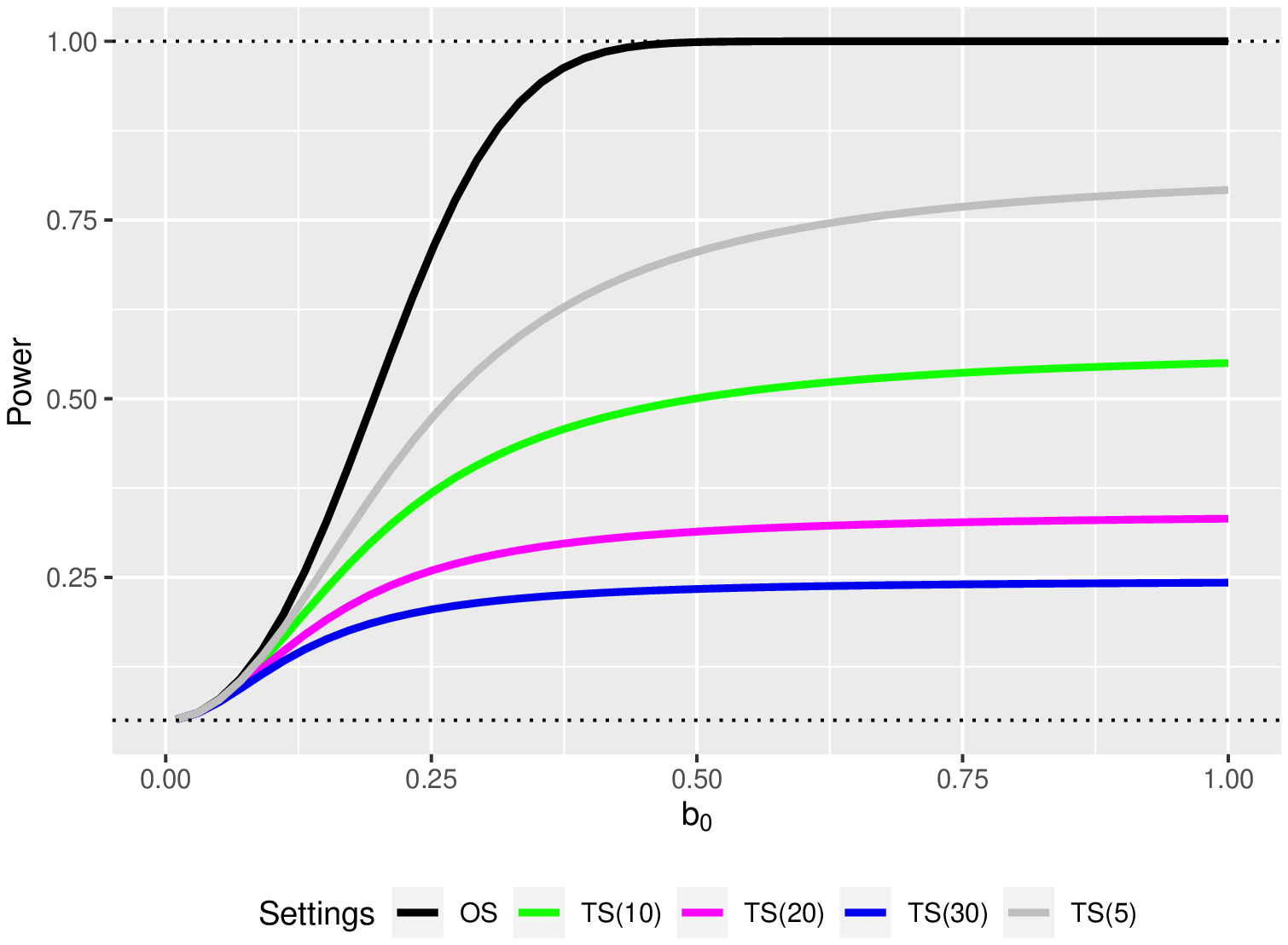}
\includegraphics[width=0.495\textwidth,height=5.7cm]{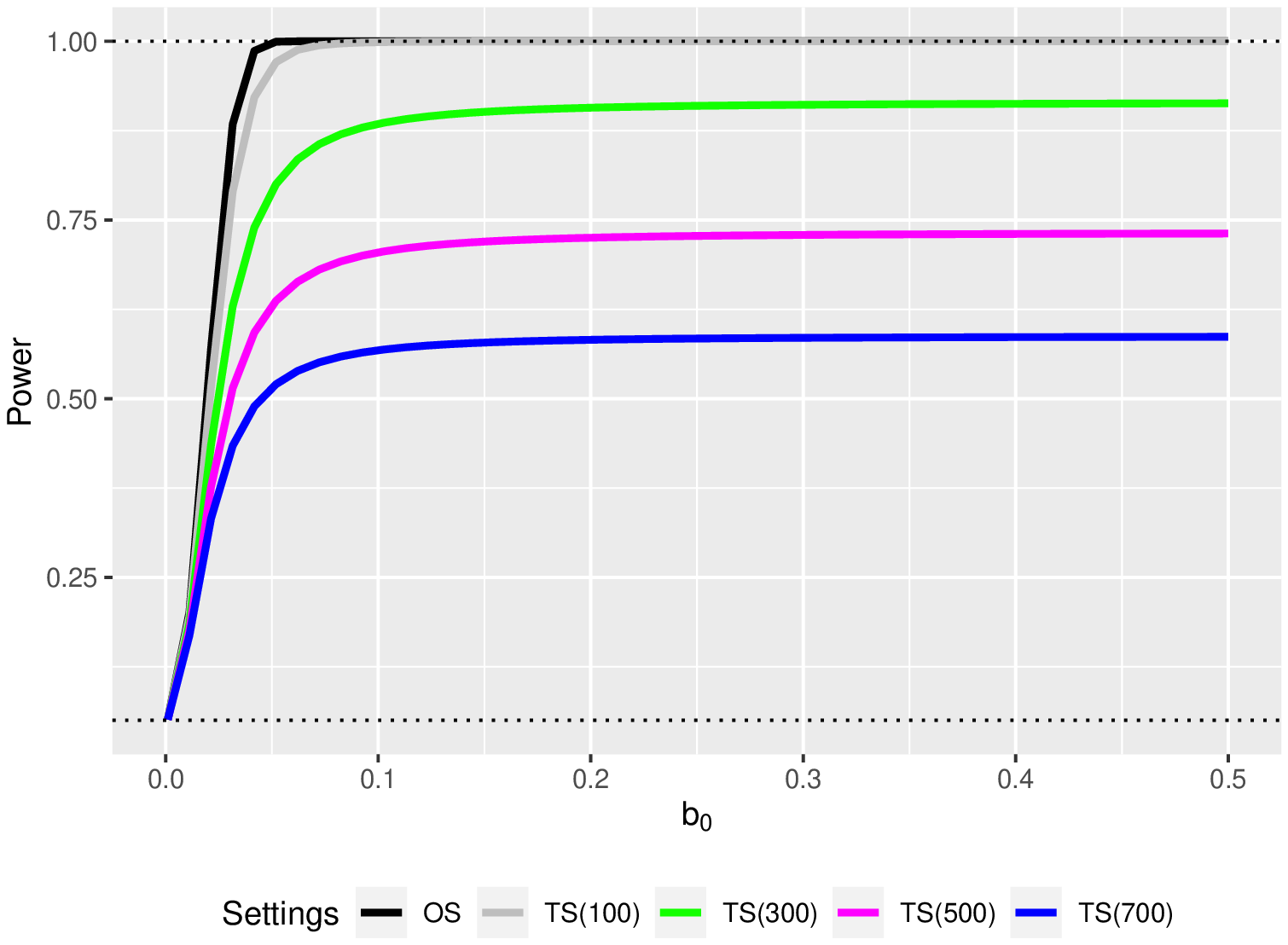}
\caption{The theoretical power functions for testing $\calH_0:\beta_j=0$. The line with ``OS'' is $\textup{Power}^{(os)}(\tau_{0.95},\bb)$ and the line with ``TS($s_0$)'' is $\textup{Power}^{(ts)}(\tau_{0.95},\bb)$ with sparsity $s=s_0$ in the equal signal strength model (\ref{em}). The left panel is $(n, \tn,p,\alpha)=(100,100,500,0.05)$ and the right panel is $(n,\tn,p,\alpha)=(10^4,5\times10^3,10^6,0.05)$. The upper dotted line indicates 1 and lower dotted line indicates the nominal level 0.05.}
\label{fig1}
\end{figure}

\begin{corollary}[Power in the equal signal strength model]
\label{cor1-st}
Consider the equal strength model (\ref{em}) with $\Sig=I_p$. The results of Theorem \ref{thm1-st} hold with
\begin{align}
\label{re1}
|\eta_j^{(os)}|=\frac{\sqrt{n}|b_0|}{\sig}~\text{and}~|\eta_j^{(ts)}|\leq\min\left\{\sqrt{\frac{n\wedge \tn}{s}},\frac{\sqrt{n}|b_0|}{\sig}\right\}.
\end{align}
\end{corollary}
In the two-sample setting, the signal strength $|\eta_j^{(ts)}|$ and hence the power is upper-truncated. As long as $|b_0|> \sig \sqrt{(n\wedge \tn)/n/s}$, the power based on proxy-data is strictly lower than the power based on one-sample data. Second, for any finite $n\wedge \tn$ and $s$, the right-hand side of  (\ref{re1}) is strictly below one no matter how large $|b_0|/\sig$ is.  This analysis demonstrates a significant loss of power in proxy-data based inference in comparison to the one-sample based inference. 
In Figure \ref{fig1}, we examine the finite sample performance based on one-sample and two-sample data in the equal signal strength model (\ref{em}). We see significant power loss with proxy data-based inference when $s$ is not too small relative to $n\wedge \tn$. 

\section{Inference for linear functionals}
\label{sec3-3}
We now  study statistical inference for the PRS $\mu_*=\bx_*^{\intercal}\bbeta$ given an individual's feature $\bx_*$. For prediction, we can use $\widehat{\mu}_*=\bx_*^{\intercal}\widehat{\bbeta}$.
We focus on constructing confidence intervals for $\mu_*=\bx_*^{\intercal}\bbeta$ in the rest of this section. Inference for linear functionals of $\bbeta$ have been studied in the classical setting. The minimax rate is established in \citet{CG17} and various methods are established in \citet{CG17}, \citet{cai2019individualized}, and \citet{javanmard2020flexible}. All the afore-mentioned methods consider the debiasing recipe: the correction scores are obtained by constrained minimizations, where the constraints can be directly used to upper bound the the remaining bias of the debiased estimator. Our problem is more challenging as some uncertainty coming from the unobserved covariance matrix $\widehat{\Sig}$ cannot be directly controlled based on the observed data.

Trading-off multiple sources of bias, we consider a different method. For $\tw_j$ defined in (\ref{eq-tw}), denote
\begin{align}
\label{eq-tOmega}
  \tOmega=(\tw_1,\dots,\tw_p)\in\R^{p\times p}.
\end{align}
In fact, $\tOmega$ is an estimate of the inverse covariance matrix. Our estimated $\tOmega$ is equivalent to the CLIME estimator \citep{Cai11}, which can be expressed as
\begin{align*}
 \tOmega&=\argmin_{\bOmega\in\R^{p\times p}}\|\bOmega\|_1\\
 &\text{subject to}~\|\tSig\bOmega-I_p\|_{\infty,\infty}\leq \tilde{\lam},
\end{align*}
where $\tilde{\lam}=c_1\sqrt{\log p/\tn}$ for some positive constant $c_1$.
We then obtain an initial bias-correction score $\tOmega\bx_*$. Next, we refine $\tOmega\bx_*$ to reduce the bias in the direction of $\bx_*$:
\begin{align}
 \tw_*&=\argmin_{\bw\in\R^{p}}\|\bw-\tOmega\bx_*\|_{1}\label{eq-tOmegas}\\
 &\text{subject to}~~\|\tSig\bw-\bx_*\|_{\infty}\leq \|\bx_*\|_2\tilde{\lam}.\nonumber
\end{align}
The optimization in (\ref{eq-tOmegas}) can be efficiently solved, because it equivalently computes the Dantzig selector \citep{CT07} by treating $\bw-\tOmega\bx_*$ as the target parameter.

For $\hat{\bbeta}^{(ts)}$ defined in (\ref{eq-cbeta}) and $\tw_*$ defined in (\ref{eq-tOmegas}), define the debiased estimator for $\mu_*=\bx_*^{\intercal}\bbeta$ as
 \begin{align}
 \label{eq-db2}
   \hat{\mu}_*^{(ts-db)}&=\bx_*^{\intercal}\hat{\bbeta}^{(ts)}+\tw_*^{\intercal}(\hS-\tSig \hbeta^{(ts)}).
\end{align}

Some more comments on $\tw_*$ defined in (\ref{eq-tOmegas}) are warranted.  Our proposed $\tw_*$ distinguishes from the constrained minimizations based on one-sample individual data, say, expressions (7) and (8) of \citet{cai2019individualized} or expression (12) of \citet{javanmard2020flexible}, directly control the bias in the direction of $\bx_*$ based on the observed $\hSig$. Their correction scores have no sparse guarantees and their objective functions are quadratic. In the proxy setting, the analysis for debiasing has more remainder terms to control, which involve the discrepancy between the observed $\tSig$ and the unobserved $\hSig$. To control the bias term involving $\hSig$, we rely on the sparsity of the precision matrix $\bOmega$. The estimator $\tw_*$ in (\ref{eq-tOmegas}) can simultaneously leverage the sparsity structure of $\bOmega$ and control the bias in the direction of $\bx_*$. 

The estimators established in Section \ref{sec3-3} can be understood as a generalization of the estimators in Section \ref{sec3}. Especially, the canonical basis ${\bm e}_j$ in (\ref{eq-tw}) is replaced by a generic linear coefficient $\bx_*$ in (\ref{eq-tOmegas}).  

\subsection{Asymptotic normality for two-sample debiased $\mu_*$}
\label{sec3-4}
We study the theoretical properties  of $\hat{\mu}_*^{(ts-db)}$ defined in (\ref{eq-db2}). The theoretical analysis for a generic linear functional is more challenging, because $\bx_*$ is non-sparse in general while canonical basis $\bm e_j$ has sparsity one. Let $s_{\Omega}=\max_{j\leq p}\|\bOmega_{.,j}\|_0$.

\begin{theorem}[Asymptotic normality of $\hat{\mu}_*^{(ts-db)}$]
\label{thm-db2}
 Assume Condition \ref{cond1} and Condition \ref{cond2} hold true, $n\gg \log p$ and $\tn\gg (s\vee s_{\Omega})\log p$.  It holds that
    \[
       \hat{\mu}_*^{(ts-db)}-\mu_*=z_*+O_P\left(\gam_{n,\tn}^{1/2}\|\bx_*\|_2\frac{(s+s_{\Omega}^{3/2})\log p}{\sqrt{n\tn}}\right),
    \]
    where $\gam_{n,\tn}$ is defined in (\ref{eq-gam}) and
    \[
       (V_*^{(ts)})^{-1/2}z_*\xrightarrow{D} N(0,1)~~ \text{for}~~ V_*^{(ts)}=\frac{\bx_*^{\intercal}\bOmega\bx_*\gam_{n,\tn}}{n}+\frac{\mu_*^2}{n}+\frac{\mu_*^2}{\tn}.
       \]
       Hence, given that \begin{align}
\label{ss2}
        s\log p \ll \sqrt{\tn}~~\text{and}~~s_{\Omega}^{3/2}\log p\ll \tn/\sqrt{n},
\end{align}
$ (V_*^{(ts)})^{-1/2}(\hat{\mu}_*^{(ts-db)}-\mu_*)\xrightarrow{D} N(0,1)$.
\end{theorem}
 Theorem \ref{thm-db2} establishes the limiting distribution of $ \hat{\mu}_*^{(ts-db)}$ and the asymptotic normality  for $\hat{\mu}_*^{(ts-db)}$ given (\ref{ss2}).
The sparsity condition on $s$ is the same as in Section \ref{sec3-2} but the condition on $s_{\Omega}$ is stricter. This comes from the challenge of dealing with a non-sparse loading $\bx_*$. 
We now connect Theorem \ref{thm-db2} with the method in (\ref{eq-tOmegas}). We see that the remaining bias of $\hat{\mu}_*^{(ts-db)}$ depends on the sparsity of $\bOmega$, $s_{\Omega}$. We leverage the sparsity of $\bOmega$ by first initialize $\tOmega\bx_*$ and compute $\tw_*$ as its projection to the $\ell_{\infty}$-constrained space. Again, the number of proxy data $\tn$ determines the range of sparsity condition for constructing confidence intervals. The asymptotic variance $V_*^{(ts)}$ is determined by $n$ and $\tn$ simultaneously.

We introduce the variance estimator of $\hat{\mu}_*^{(ts-db)}$, which is
\begin{align}
   \widehat{V}^{(ts)}_*&=\tw_*^{\intercal}\tSig\tw_*(\frac{\|\by\|_2^2}{n^2}+\frac{2(\hbeta^{(ts)})^{\intercal}\hS-(\hbeta^{(ts)})^{\intercal}\tSig\hbeta^{(ts)}}{\tn})\nonumber\\
   &\quad+\frac{(\hat{\mu}_*^{(ts-db)})^2}{n}+\frac{(\hat{\mu}_*^{(ts-db)})^2}{\tn}.\label{eq-Vnew}
\end{align}
We can similarly show that $\widehat{ V}_*^{(ts)}$ defined in (\ref{eq-Vnew}) is a consistent estimator of $V_*^{(ts)}$. Hence, we propose the following $(1-\alpha)\times 100\% $-confidence interval for $\mu_*$ as
\begin{equation}
\label{ci2}
    \hat{\mu}_*^{(ts-db)}\pm \tau_{\alpha/2}\sqrt{  \widehat{V}^{(ts)}_*}.
\end{equation}

\section{Numerical results}
\label{sec-simu}
In this section, we evaluate the empirical performance of the estimation and inference procedures developed in previous sections.
A practical issue is the choice of tuning parameter $\lam^{(ts)}$. Without the individual data, cross-validation cannot be used. Alternative strategies include using some information criteria including Akaike information criterion (AIC) and Bayesian information criterion (BIC), or generalized information criterion. Some theoretical guarantees based on theses criteria have been studied, see, for example, \citet{zhang2010regularization} and \citet{fan2013tuning}. Here we use BIC to select $\lam^{(ts)}$ in (\ref{eq-cbeta}). Specifically, we consider
\[
  \textup{BIC}(\lam)= \log\left((\hat{\bbeta}^{(ts)}_{\lam})^{\intercal}\tSig\hat{\bbeta}^{(ts)}_{\lam}- 2(\hat{\bbeta}^{(ts)}_{\lam})^{\intercal}\hS+M\right)+\frac{\log (n\wedge \tn)}{n\wedge \tn} \|\hat{\bbeta}^{(ts)}_{\lam}\|_0,
\]
where $\hat{\bbeta}^{(ts)}_{\lam}$ is the two-sample Lasso estimate with tuning parameter $\lam$.
The variance of $\hat{\beta}_j^{(ts-db)}$ depends on $M$, the second moment of $y_i$. In fact, $M$ can be approximated from the variance estimators of $\hS$. Specifically, it is easy to show that 
\[
   \textup{Var}(\hS_j)=\textup{Var}\left(\frac{\bX_j^{\intercal}\by}{n\Sig_{j,j}}\right)(1+o(1))=\frac{\Sig_{j,j}M+(\Sig_{j,S}\bbeta_S)^2}{n\Sig_{j,j}^2}(1+o(1)).
\]
When the correlation between $\bx_{.,j}$ $j\notin S$ and $\bX_{.,S}$ is sparse, say, $|\{j\leq p: \Sig_{j,S}\neq 0\}|\ll p$, then a consistent estimate of $M$ is
\[
   \widehat{M}=\frac{n}{p}\sum_{j=1}^p \widehat{var}(\hS_j)\tSig_{j,j}.
\] 

For one-sample method, the tuning parameter in Lasso is chosen based on 10-fold cross validation and the tuning parameter in debiased Lasso is set as $\sqrt{2\log p/n}$.

 We consider $p=500$ and $s\in\{4,8,12\}$.
Let $\beta_{4(k-1)+1: 4k}=(0.5,-0.5,0.2,-0.2)^{\intercal}$ for $k=1,\dots, s/4$ and $\beta_k=0$ otherwise.
We consider $(n,\tn)\in\{(100,400),$ $(200,400),(200,200),(400,200),(400,100)\}$, which corresponds to $n/\tn\in\{0.25,0.5,1,2,4\}$. In the main paper, we consider the identity covariance matrix and an equi-correlated $\Sig$ with $\Sig_{j,k}=1$ if $j=k$ and $\Sig_{j,k}=0.3$ if $j\neq k$. In the Supplemental Materials, we also provide results based on a block diagonal $\Sig$ where each block is Toeplitz with the first row being $(0.6^0,0.6,\dots, 0.6^4)$.

\subsection{Estimation and prediction results}
\label{sec-simu-1}

\begin{figure}[H]
\includegraphics[width=0.99\textwidth, height=4.7cm]{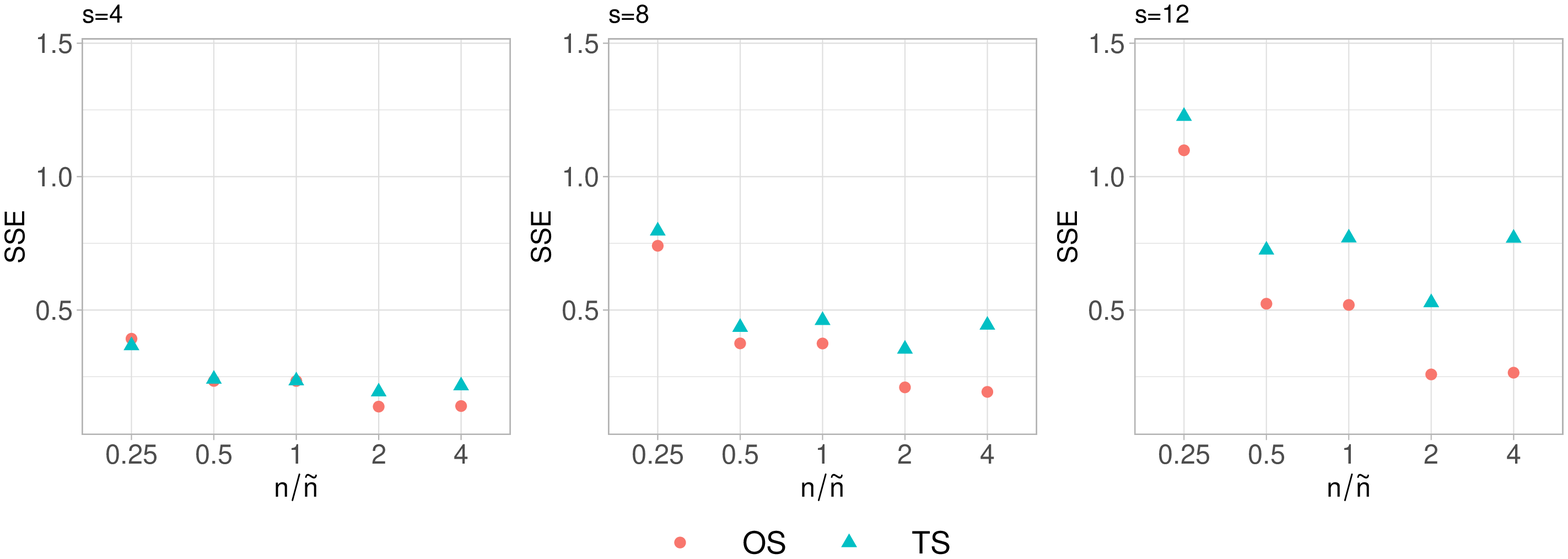}
\includegraphics[width=0.99\textwidth, height=4.7cm]{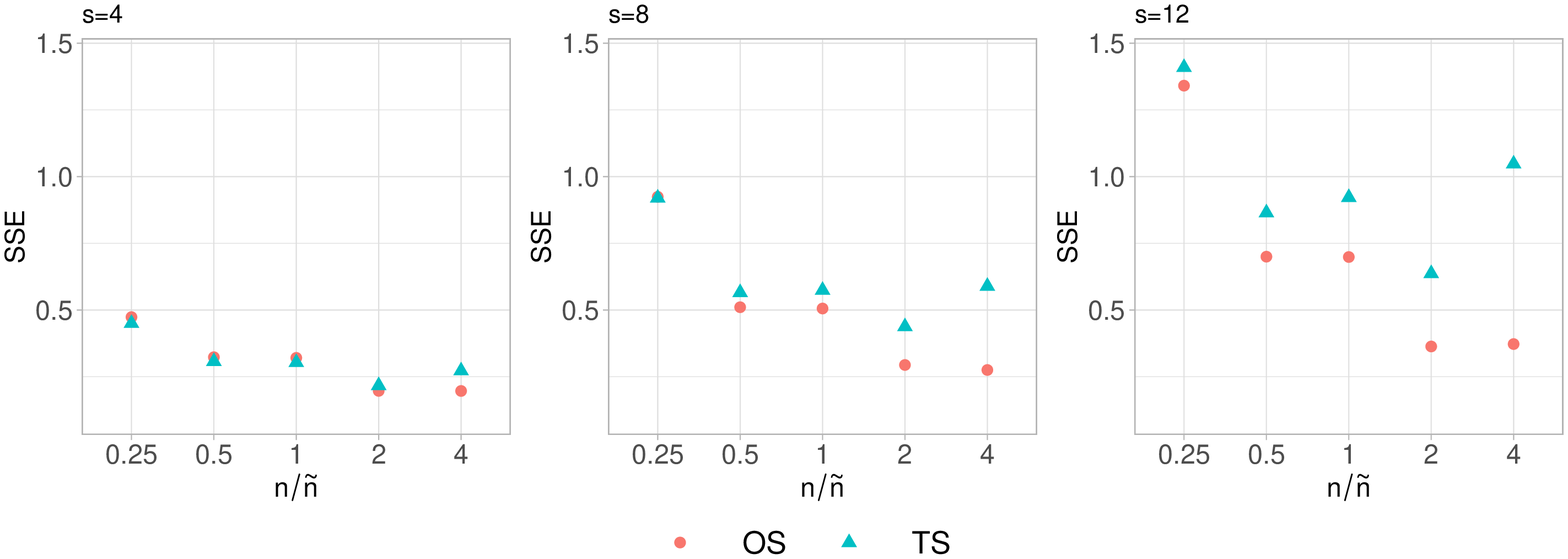}
\caption{\label{fig-SSE}Sum of squared errors (SSE) for estimating $\bbeta$ based on one-sample Lasso (OS) and two-sample Lasso (TS) with identify covariance matrix (first row) and equi-correlated covariance matrix (second row). Each point is the mean based on 200 independent experiments. }
\end{figure}

In Figure \ref{fig-SSE}, we report the estimation results with one-sample  and two-sample Lasso. The prediction results are similar and are given in the supplements. As one-sample method only uses $n$ individual samples, we see that the estimation errors (SSE) decrease as $n$ increases for any given $s$. 
For the methods based on proxy data, the estimation errors do not monotonically decrease as $n/\tn$ increase. This is because the estimation error is proportional to $\gam_{n,\tn}/n$, which depends on $n\wedge \tn$. Second, the estimation errors are larger than the corresponding errors in one-sample case. As the sparsity $s$ increases, both methods have SSE increasing. However, the SSE of two-sample Lasso increases more significantly. This is because, according to Theorem \ref{lem2-est}, as $s$ increases, SNR increases and a larger $\lam^{(ts)}$ is needed which leads to larger SSE. In the conventional one-sample case, the tuning parameter $\lam^{(os)}$ can be chosen independent of SNR theoretically.
For equi-correlated $\Sig$, we observe similar patterns on estimation errors as for $\Sig=I_p$. 

\subsection{Inference for the regression coefficients}
\label{sec-simu-2}
We evaluate the performance of different methods for statistical inference for the regression coefficients. Besides the one-sample and two-sample debiased Lasso, we also evaluate the effects of treating the two-sample summary data as one-sample data on statistical inference. We have discussed in Section \ref{sec3} that it can lead to incorrect variance estimators for the debiased Lasso. Hence, we consider a method with misspecification, which applies one-sample formula to two-sample data, shorthanded as TS2.

\begin{figure}[H]
\includegraphics[width=0.99\textwidth, height=6cm]{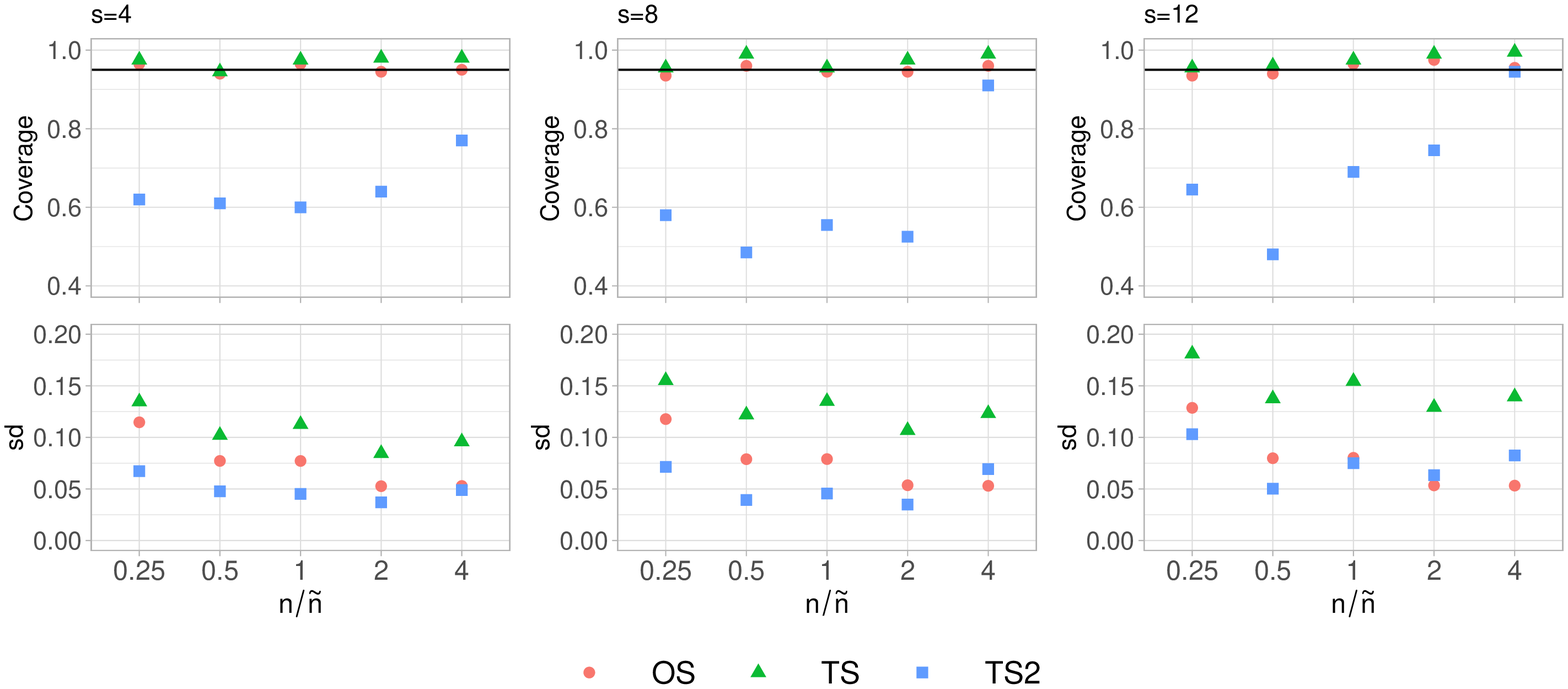}
\caption{\label{fig-ident-coef0}Average coverage probabilities (first row) and average standard deviations (second row) with identify covariance matrix for $\beta_{20}=0$. Three method in comparison are one-sample debiased Lasso (OS), two-sample debiased Lasso (TS), and the application of one-sample debiased Lasso to two-sample data (TS2). The solid line is the nominal confidence level 0.95. Each point is the mean based on 200 independent experiments.}
\end{figure}

In Figure \ref{fig-ident-coef0}, we see that the one-sample and two-sample debiased Lasso for a zero coefficient has coverage probabilities close to the nominal level in various configurations. The misspecified method, however, has coverage probabilities significantly lower than the nominal level. These results highlight the importance of dealing with proxy data with the methods introduced in Section \ref{sec-method}. Ignoring the fact that the covariance matrix is estimated from the proxy data can lead to incorrect statistical inference. On the other hand, the standard deviations based on proxy data are significantly larger than those based on one-sample data especially when $n$ is large and $\tn$ is small. The gap increases as SNR increases. This agrees with our analysis in Theorem \ref{thm-db}. We also point out that in contrast to the one-sample case, the standard deviations in two-sample case are not monotonic functions of $n/\tn$.

\begin{figure}[H]
\includegraphics[width=0.99\textwidth, height=6cm]{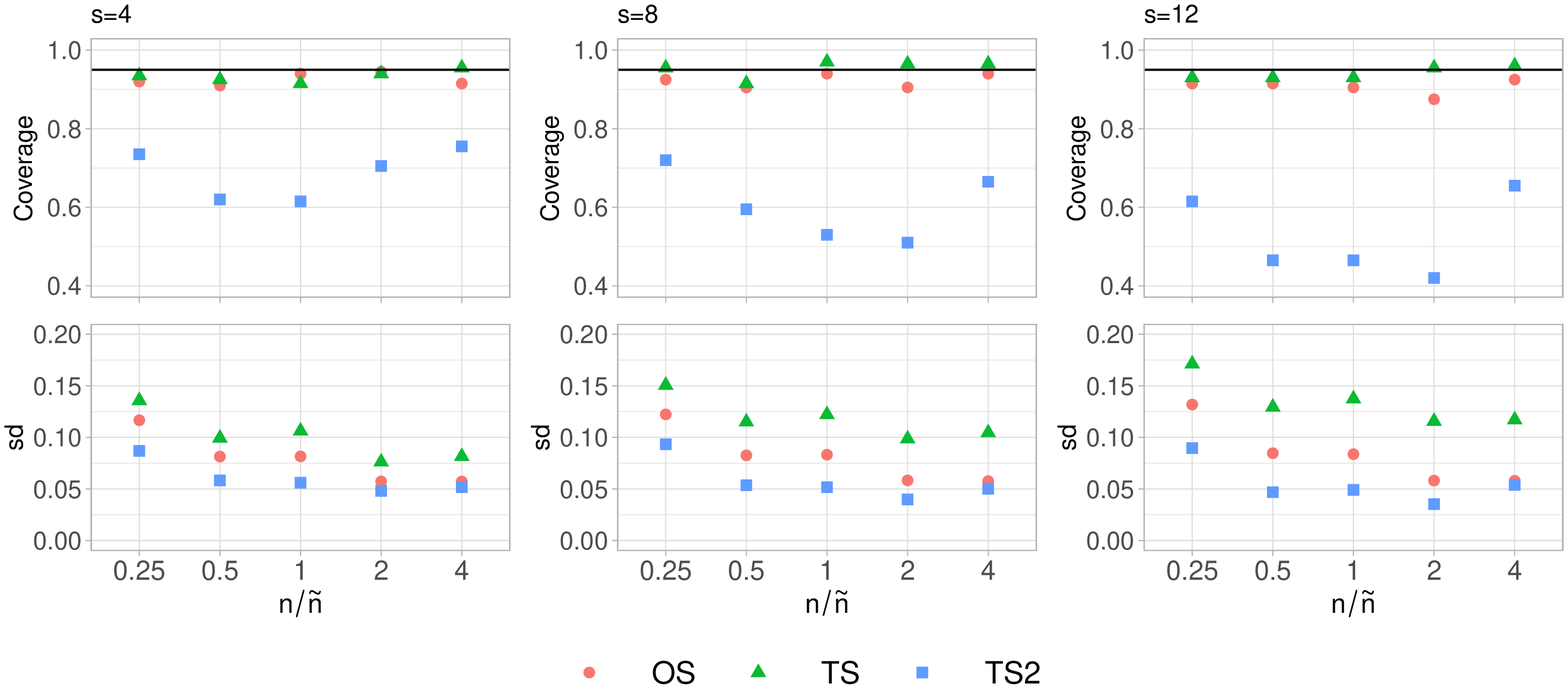}
\caption{\label{fig-equi-coef1}Average coverage probabilities (first row) and average standard deviations (second row) with equi-correlated matrix for $\beta_3=0.2$. Three method in comparison are one-sample debiased Lasso (OS), two-sample debiased Lasso (TS), and the application of one-sample debiased Lasso to two-sample data (TS2). The solid line is the nominal confidence level 0.95. Each point is the mean based on 200 independent experiments. }
\end{figure}

In Figure \ref{fig-equi-coef1}, we report the one-sample and two-sample debiased Lasso for $\beta_{3}=0.2$ with non-sparse $\bOmega$.
The coverage probabilities close to the nominal level in most configurations but are slightly lower when $s$ is large. This is mainly due to a non-sparse $\bOmega$ can cause larger remaining bias in the debiased Lasso method. 

The results for $\beta_1=0.5$ are provided in the Supplemental Materials. We see that for strong signals, the coverage probabilities are slightly lower than the nominal level in the equi-correlated setting and block-diagonal setting. This is because the debiased estimators for strong signals are subject to larger remaining bias and similar patterns have been observed and studied in one-sample setting. In all the settings, we see that the coverage probabilities for $\beta_{20}=0$ are close to the nominal level. Hence, testing $H_{0,j}:\beta_j=0$ can always have accurate Type-I error in all the scenarios in consideration.

\subsection{Inference for the linear functionals}
\label{sec-simu-3}
\begin{figure}[H]
\includegraphics[width=0.99\textwidth, height=6cm]{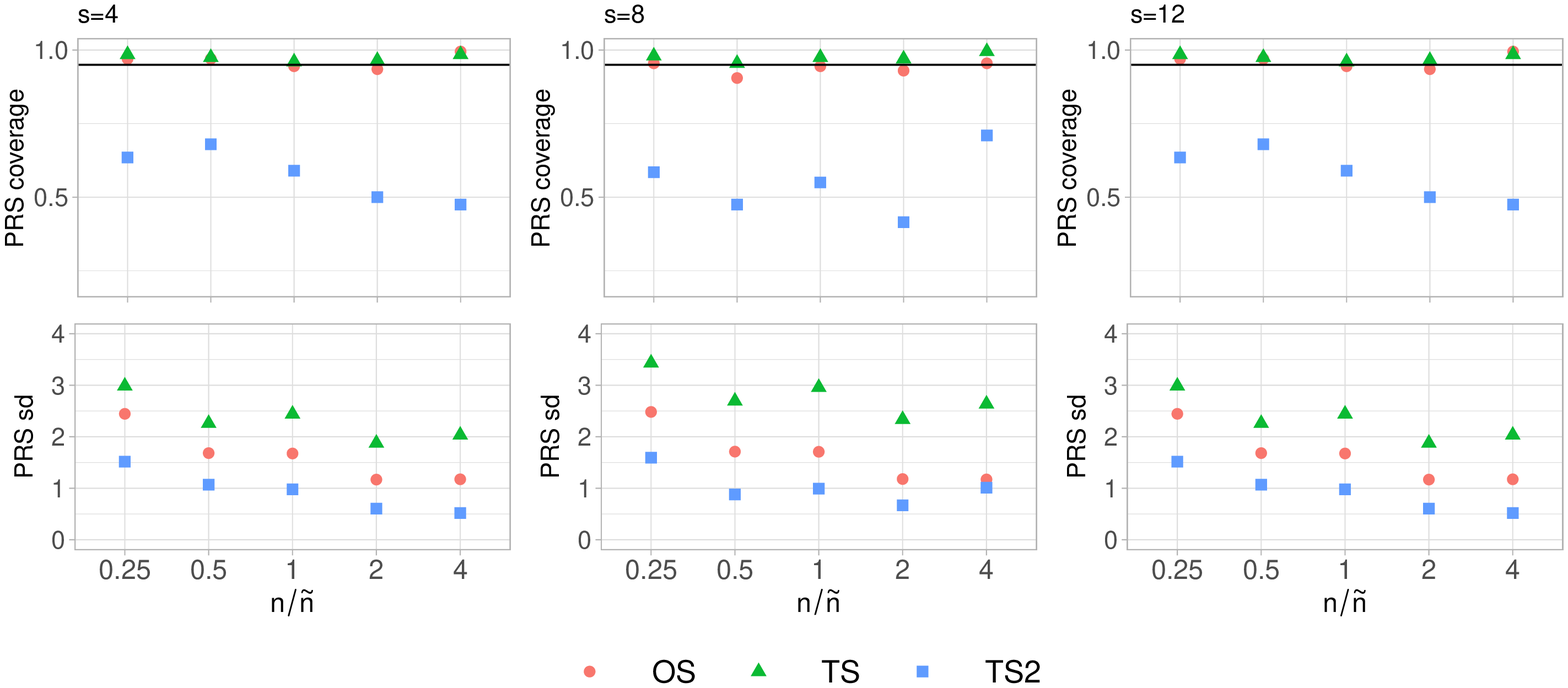}
\caption{\label{fig-mu-ident}Average coverage probabilities (first row) and average standard deviations (second row) with identify covariance matrix for $\mu_*$. Three method in comparison are one-sample debiased Lasso (OS), two-sample debiased Lasso (TS), and the application of one-sample debiased Lasso to two-sample data (TS2). The solid line is the nominal confidence level 0.95. Each point is the mean based on 200 independent experiments. }
\end{figure}

\begin{figure}[H]
\includegraphics[width=0.99\textwidth, height=6cm]{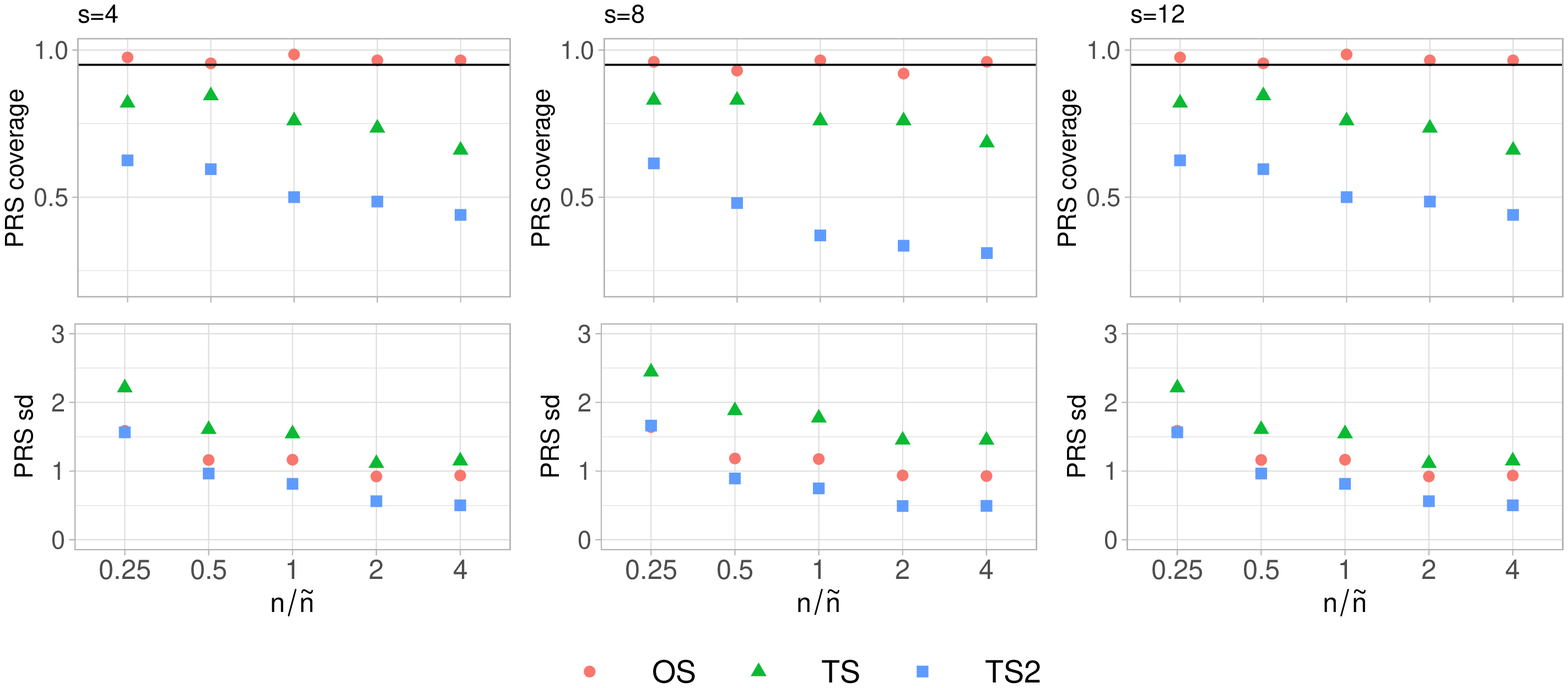}
\caption{\label{fig-mu-equi}Average coverage probabilities (first row) and average standard deviations (second row) with equi-correlated matrix for $\mu_*$. Three method in comparison are one-sample debiased Lasso (OS), two-sample debiased Lasso (TS), and the application of one-sample debiased Lasso to two-sample data (TS2). The solid line is the nominal confidence level 0.95. Each point is the mean based on 200 independent experiments. }
\end{figure}

In this subsection, we present the inference results for $\bx_*^{\intercal}\bbeta$ where $\bx_*$ is randomly generated from $N(0,\Sig)$ and $\Sig$ is taken to be the identity or the equi-correlated matrix, respectively.  With $\Sig=I_p$ (Figure \ref{fig-mu-ident}), the coverage probabilities of one-sample and two-sample methods have coverage probabilities close to the nominal level. However, mis-fitting two-sample data with one-sample method still lead to low coverage For non-sparse $\bOmega$ (Figure \ref{fig-mu-equi}), the two-sample coverage probabilities are lower than the nominal level. For block diagonal $\Sig$, which corresponds to a sparse $\bOmega$, the coverage probabilities are close to the nominal level. These observations reveal that the two-sample methods are more severely affected by the sparsity of precision matrix.
Moreover, the coverage gets lower as $n/\tn$ increases. This again witnesses the critical role of $\tn$ for two-sample inference. See the sample size condition (\ref{ss2}) and the discussion follows.

\section{Data analysis}
\label{sec-data}
We apply the proposed methods to a GWAS study in outbred Carworth Farms  White (CFW) mice population \citep{parker2016genome}. \cite{parker2016genome} showed no widespread population structure or cryptic relatedness in the CFW mice and therefore, we view these mice as independent of each other. The primary pre-processing of phenotypes and genotypes, including outliers removal and basic transformation, was conducted using the original paper's code. After the pre-processing, the data set consists of 1,038 mice with 79,824 genetic variants (SNPs) and 71 different phenotypes.  We study the genetic associations for the weights of four hindlimb muscles. Specifically, the responses include the weight of TA (transverse abdominal), EDL(extensor digitorum longus), gastroc (gastrocnemius), and soleus, respectively.  

\subsection{Prediction of hindlimb muscle weights using genotype data}
In prediction tasks, we use all the SNPs to predict each response and evaluate the out-of-sample prediction accuracy. Take the TA response as an example. In each experiment, we randomly split the samples into two folds and use one fold to compute the GWAS statistic, $\hS$, and the other fold to compute the sample covariance matrix, $\tSig$. We consider different sample size ratios of the GWAS and the empirical covariance matrix. Specifically, we consider the sample size for GWAS, $n\in\{250,500,750\}$, and $\tn=1038-n$, which gives $n/\tn$ is approximated one of $\{1/3,1,3\}$. For each sample size configuration, we repeat independent splitting and predictions 30 times.

\begin{figure}[H]
\includegraphics[height=5.2cm,width=0.99\textwidth]{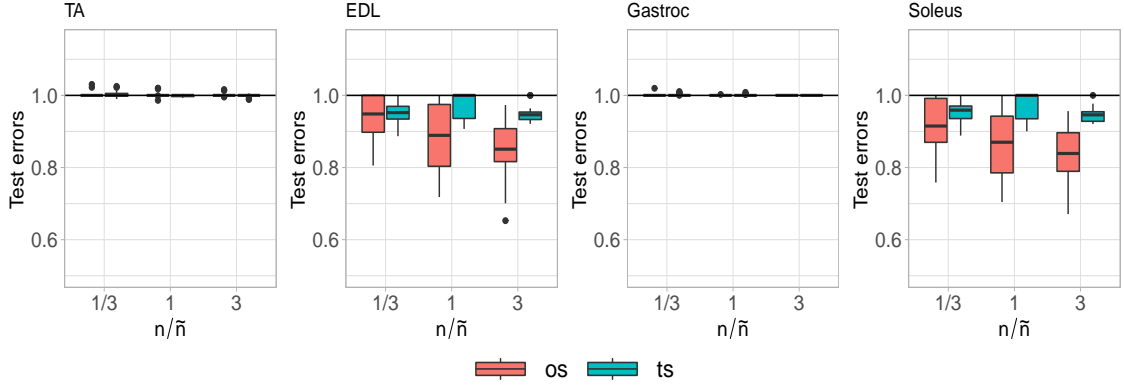}
\caption{Test errors based on one-sample and two-sample Lasso prediction for four muscle weights. The x-axis denotes three settings corresponding to $n/\tn\in\{1/3,1,3\}$ and y-axis reports the relative test error $\|\tilde{\by}-\tilde{\bx}^{\intercal}\bm{b}\|_2^2/\|\tilde{\by}\|_2^2$. Each boxplot is based on 30 random splits.}
\label{CFW-pred}
\end{figure}

The prediction results are plotted in Figure \ref{CFW-pred}. We observe  that the SNPs are predictive for the  EDL and Soleus weight, but are not predictive for the TA and Gastroc weight  for any sample size ratios. For EDL and Soleus, we see that as $n/\tn$ in creases, the test errors decrease significantly in one-sample case. 
In two-sample case, the test errors have the smallest median when $n/\tn=1$. This can be understood through a simple analysis. For a fixed total sample size $n+\tn=N$ and some $0<\rho<1$, the term 
\[
  \gam_{n,\tn}/n=\gam_{\rho N,(1-\rho)N}/(\rho N)=\frac{M-\rho\sig^2}{N\rho(1-\rho)}.
  \] 
  Approximating the numerator by $cM$ for some $0<c<1$, it gives that $\rho=1/2$ minimizes $\gam_{\rho N,(1-\rho)N}$. Recall that $\gam_{n,\tn}/n$ determines the convergence rate of the two-sample Lasso as shown in Theorem \ref{lem2-est}. This explains why $n/\tn=1$ has the smallest test errors in two-sample case. Comparing the one-sample and two-sample results, we see that one-sample Lasso has more accurate predictions on average for different sample size ratios.

\subsection{Inference for polygenic risk scores}
\label{sec6-3}
In order to build a polygenic risk prediction,  we first perform a pre-processing step to remove highly correlated covariates and to reduce the computational cost. We divide all the SNPs along chromosomes 1 to 19 into five blocks with  each block containing about 15,000 SNPs. We perform principle component analysis to each block so that the principal components (PCs) account for  90\% of variation of the SNPs of that block.
This gives 3,302 principal components and we use them as the design matrix. We mention that the PCs of SNPs are linear transformations of SNPs and hence the corresponding regression coefficients are also linear transformations of the original coefficients. However, it is easy to see that the definition of $\mu_*$ is invariant to the linear transformations on the designs. 

\begin{figure}[H]
\includegraphics[height=6.5cm,width=0.99\textwidth]{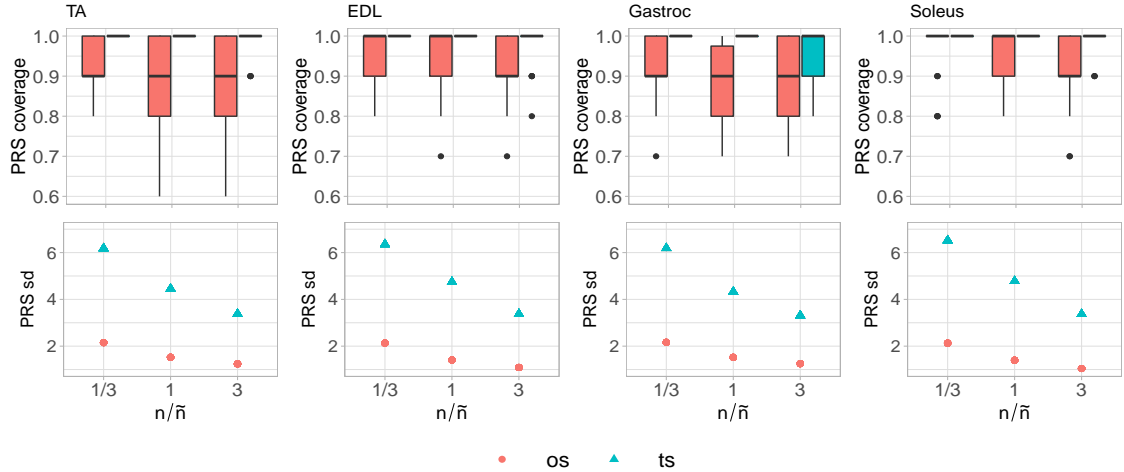}
\caption{Inference for PRS based on one-sample and two-sample debiased Lasso for four muscle weights. The x-axis denotes three settings corresponding to $n/\tn\in\{1/3,1,3\}$ and y-axis reports the average probabilities of the 95\% confidence intervals covering $\tilde{y}_*$ (top) and the average standard errors (bottom) in each experiment. Each boxplot is based on 30 random splits.}
\label{CFW-PRS}
\end{figure}

 In each experiment, we randomly select 10 pairs of $(\bx_*,\tilde{y}_*)$ from $\{(\tx_i,\tilde{y}_i)\}_{i=1}^{\tn}$ and report the average coverage of proposed confidence intervals on $\tilde{y}_*$. The results are given in Figure \ref{CFW-PRS}. We mention that as $\tilde{y}_*\neq \mu_*$ and $\tilde{y}_*$ has larger variance from noise, the coverage probabilities can be lower than the nominal level. 
We see that the one-sample method has median coverage of above 90\% in all the settings. The two-sample method has coverage probabilities close to one in most settings.
In view of the lengths of the confidence intervals, we see that the one-sample method results in confidence intervals that are significantly shorter than those of two-sample method.

Finally, we present in the Supplemental Materials the inference results for the individual coefficients. We observe that the confidence intervals based on one-sample data are much shorter than those based on proxy data, further indicating the power limit for proxy-data-based inference.

\section{Discussion}
\label{sec-diss}
Statistical learning with summary data has attracted  significant interests in genetic, epidemiology, and other health-related studies.
In this work, we have provided statistical inference methods and theoretical guarantees with proxy data in high-dimensional linear models. Some new phenomenon are observed in the asymptotic normality conditions and power functions.

We emphasize that the challenges stem  from both the proxy and summary properties of the available data.  One  challenge with summary data-based regression is on tuning parameter selection and we consider BIC in Section \ref{sec-simu} and discuss some other options. Another  challenge emerges from  existence of heteroscedastic noises and model misspecification. In these two cases, the asymptotic variance of the debiased Lasso need to be estimated empirically \citep{buhlmann2015high,Dezeure17}, which cannot be applied with summary data. These are important and interesting future research directions.

We conclude by pointing out some other related proxy-data problems. First, it is interesting to study classification based on the proxy data. The corresponding models, such as the logistic regression model, have nonlinear link functions and hence the external covariance matrix cannot be directly used as a proxy for the Hessian. Further approximations are needed.  Second, in genetic studies, LD-score regression \citep{bulik2015ld,speed2019sumher} based on GWAS summary statistics has been widely applied to  estimate the heritability and co-heritability, where   linear mixed-effects models are assumed. Finally,  causal effect estimation using  two-sample Mendelian randomization \citep{hartwig2016two,bowden2017framework} is widely studied in epidemiology. It leverages the GWAS summary data for the exposure and for the outcome, which can be collected based on different samples, to conduct causal inference.  These  problems may exhibit similar theoretical properties  as investigated in this work but are not included in our linear model setting. It is of significant interest to study the potential bias and power loss in these problems when only summary statistics are available.

\section*{Acknowledgments}
S.L. was supported by the Fundamental Research Funds for the Central Universities, and the Research Funds of Renmin University of China.
T.T.C and H.L. were supported by NIH grants GM129781 and GM123056. 

\bibliographystyle{chicago}
\bibliography{ts.bib}

\end{document}